\documentclass[preprint,12pt]{elsarticle}
\usepackage{graphicx}     % For images
\usepackage{amssymb}      % Math symbols
\usepackage{amsmath}      % Math formatting
\usepackage{siunitx}      % SI units
\usepackage{caption}      % Captions
\usepackage{subcaption}   % Subfigures
\usepackage{bm}           % Bold math
\usepackage{float}        % Precise float placement
\usepackage{soul}         % Underlining, highlighting
\usepackage{lineno}       % Line numbers (use \linenumbers later)

\journal{--}

\begin{document}

\begin{frontmatter}

\title{Testing a Computed Tomography Imaging Spectrometer for Earth Observations on the HEIMDAL Stratospheric Balloon Mission}

\author[sduphysics,sduclimate]{Mads Juul Ahlebæk}\corref{cor1}
\ead{ahle@sdu.dk}
\author[sdu_ing]{Albertino Antonio Almeida Bach}
\author[sdu_ing]{Loui Collin-Enoch}
\author[au_ece]{Christian Cordes}
\author[au_cs]{Boas Hermansson}
\author[sduphysics]{Christian Hald Jessen}
\author[sdu_ing]{Søren Peter Jørgensen}
\author[sdu_ing]{Tobias Jørgensen}
\author[sdu_ing]{Viktor Ulrich Kanstrup}
\author[sduphysics]{Laurits Tværmose Nielsen}
\author[au_physics]{Jes Enok Steinmüller}
\author[newtec]{Mads Svanborg Peters}
\author[au_physics]{Jonathan Merrison}
\author[mci,sduclimate]{René Lynge Eriksen}
\author[au_physics]{Christoffer Karoff}
\author[sduphysics,sduclimate]{Mads Toudal Frandsen}

\cortext[cor1]{Corresponding author}

\affiliation[sduphysics]{organization={Department of Physics, Chemistry and Pharmacy, University of Southern Denmark},
            city={Odense},
            country={Denmark}}
\affiliation[sduclimate]{organization={SDU Climate Cluster, the University of Southern Denmark},
            city={Odense},
            country={Denmark}}
\affiliation[sdu_ing]{organization={Faculty of Engineering, University of Southern Denmark},
            city={Odense},
            country={Denmark}}
\affiliation[au_ece]{organization={Department of Electrical and Computer Engineering, Aarhus University},
            city={Aarhus},
            country={Denmark}}
\affiliation[au_cs]{organization={Department of Computer Science, Aarhus University},
            city={Aarhus},
            country={Denmark}}
\affiliation[au_physics]{organization={Department of Physics and Astronomy, Aarhus University},
            city={Aarhus},
            country={Denmark}}
\affiliation[mci]{SDU NanoSyd, Mads Clausen Institute, University of Southern Denmark}
\affiliation[newtec]{organization={Newtec Engineering A/S},
            city={Odense},
            country={Denmark}}

%% Abstract
\begin{abstract}
Stratospheric High Altitude Balloons (HABs) have great potential as a remote sensing platform for Earth Observations that complements orbiting satellites and low flying drones. At altitudes between 20-35 kms, HABs operate significantly closer to ground than orbiting satellites, but significantly higher than most drones. HABs therefore offer a unique potential to deliver high spatial resolution imaging with large area coverage. Another two imaging parameters that are important for Earth Observation applications are spectral resolution and spectral range. In this paper, we therefore present the development and testing of a hyperspectral imaging system, able to record near-video-rate images in narrow contiguous spectral bands, from a HAB platform. 
In particular, we present the first stratospheric environmental tests and HAB flight of a snapshot hyperspectral camera, based on Computed Tomography Imaging Spectroscopy (CTIS), which is well suited to cope with the challenges posed by the motion of the HAB platform and the stratospheric environment. We have successfully acquired images with the system under both simulated stratospheric conditions in the Mars Simulation Laboratory at Aarhus University and during a 5 hour HAB flight mission named HEIMDAL from Kiruna in October 2024 as part of the REXUS/BEXUS 34/35 2024 campaign organized by DLR-SNSA. The study represents a step towards deploying the HAB platform for high quality land cover classification. 
\end{abstract}

%% Keywords

\end{frontmatter}

\section{Introduction}
Stratospheric High Altitude Balloons (HABs) offer a reusable and adaptable platform for remote sensing that can accelerate the availability of high resolution Earth Observation (EO) data~\citep{Floberghagen}. At an altitude of 20-35 kms they operate a factor of 20 closer to the surface of the earth than Low Earth Orbit Satellites. This allows HABs to acquire higher resolution optical imagery than satellites with comparable optics, while still covering significantly larger areas, and stay afloat for longer times, than conventional drones. HABs therefore have a very significant potential as an EO platform~\citep{Pankine} --- both in themselves and in co-operation with satellites to deliver both very large area coverage and very high spatial and temporal resolution EO~\citep{Drinkwater,Floberghagen}.

Two other important imaging parameters for EO imaging is spectral resolution and range. Therefore HABs should also be able to carry payloads for spectral imaging --- a hybrid modality that combines imaging and spectroscopy with spectral range and resolution beyond Red-Green-Blue (RGB).    
Hyperspectral Imaging (HSI) \citep{AlexanderF} is a specific spectral imaging technique that combines imaging and spectroscopy by recording light intensity in many narrow contiguous spectral bands at each spatial point $(x,y)$. Since different features on Earth’s surface, such as trees or water bodies, reflect (absorb, scatter and emit) light differently at specific wavelengths, this creates a characteristic spectral fingerprint that can be identified with HSI.

A main HSI acquisition technique is pushbroom (spatial) scanning \citep{boldrini2012hyperspectral} where the spectral information of a one spatial dimension slice is projected onto the sensor. But pushbroom scanning requires the velocity of the imaged objects relative to the camera to be known to high precision. Uncertainties in the relative motion can result in significant motion artifact and distortions of the image. The pushbroom technique is therefore well suited to large satellites with little motion relative to the orbit, but less well suited for HABs where the gondola in general is moving more relative to the flight trajectory. 

In this paper we therefore study a snapshot (non-scanning) HSI technique where the whole scene is recorded in a single integration time of the sensor.  This reduces sensitivity to motion artifacts from the HAB platform. The particular snapshot HSI system we develop and test is a computed tomography imaging spectrometer (CTIS) \citep{descour_computed-tomography_1995,Th,Okamoto:91}, developed by Newtec Engineering A/S. The imaging system was developed and tests flown as part of the HEIMDAL mission. The test flight launched from Kiruna in October 2024 as part of the REXUS/BEXUS 34/35 2024 campaign organized by DLR-SNSA.

The prototype camera has a spectral range from 600-850 nm and captures spatial and spectral information simultaneously by utilizing a 2D diffraction grating to disperse the light into a $3\times3$ diffraction pattern of a central zeroth order and 8 surrounding first orders. From this diffraction image, a tomographic reconstruction is required to obtain a hyperspectral image \citep{descour_computed-tomography_1995,Th,Okamoto:91}. The image processing system is described in~\citep{huang_application_2022,ahlebaek_hybrid_2023} and some of the first proximal sensing applications have been documented in~\citep{10.1117/12.2621128,PETERS2025126017}. 

HSI remote sensing finds many applications in monitoring of the climate, the environment and nature. Satellite- and drone-based HSI has, for example, been used extensively for land cover and biomass classification, including tree species classification~\citep{MAYRA2021112322,rs8060445,Masaitis}, as well as for studying vegetation dynamics, including in the Arctic region \citep{nelson2022}. As a simple proof of principle application of the acquired CTIS data we perform an analysis to distinguish land and water bodies from the flight.

The outline of the paper is as follows: In section~\ref{sec:CTISSystem} we present the CTIS HAB imaging system and image reconstruction algorithmss. In section~\ref{Imaging System Integration} we discuss the integration of the imaging system into the BEXUS HAB gondola, the mission control software and lab tests. In section~\ref{sec:operations} we discuss the actual BEXUS HAB flight launched from Kiruna and in section~\ref{sec:results} we present an analysis of the acquired data.

\section{HEIMDAL HAB Imaging System}\label{sec:CTISSystem}
The technical objective of the mission presented here is to successfully acquire images with the CTIS system under emulated stratospheric conditions and during flight. 

Despite the potential of HABs for providing high resolution EO imaging --- both spatially and spectrally --- there is only little literature on HSI imaging systems for HABs~\citep{Delaure,navarro_novel_2022}.  
To our knowledge, the study presented here is the first to demonstrate a CTIS imaging system on a HAB and one of the first to demonstrate snapshot hyperspectral imaging from a HAB.

\subsection{The Imaging system}\label{subsec:setup}
\begin{figure}[H]
\centering
\includegraphics[width=0.8\linewidth]{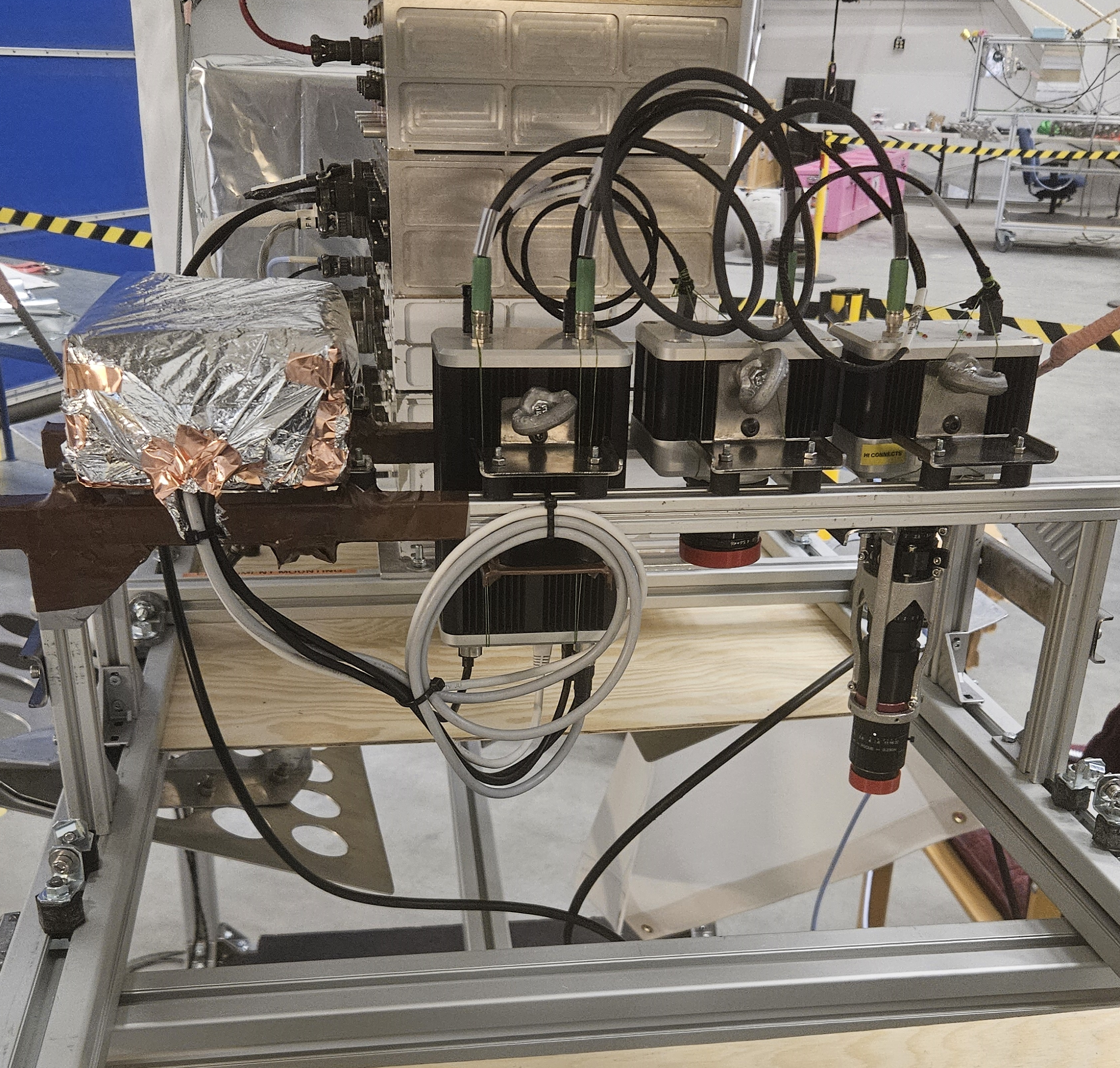}
\caption{Picture of the CTIS imaging system mounted in the BEXUS Gondola before launch}
\label{fig:AssemblyLaunchCampaign}
\end{figure}

The complete imaging system is shown integrated into the HAB Gondola in Figure~\ref{fig:AssemblyLaunchCampaign}. 

The four modules are, from right to left, the CTIS camera (long lens with red cap), a companion RGB Camera, the on board computer (OBC) and the Electronics box.

The camera system (OBC, RGB Camera and CTIS Camera) is co-developed by Newtec Engineering A/S and Qtechnology A/S, and is based on their QT-5222 platform. This system is a dual-head camera system, meaning that both cameras are controlled, triggered, and all acquired images are handled by the OBC. It is an upgraded version of the single-head system described in~\citep{peters_high-resolution_2022} and~\citep{PETERS2025126017}, where the OBC and the CTIS camera were combined into a single module. Despite the change in system architecture, the CTIS optics and specifications remain the same: it houses a monochrome 4 MP GSENSE2020 CMOS sensor, a custom diffractive optical element (DOE), two 35 mm Vis-NIR VS-technology lenses (VS-H3520-IRC), and an outer 50 mm Vis-NIR VS-technology lens (VS-H1620-IRC), effectively determining the focal length of the camera. The spectral range of the camera is 600–850 nm, defined by a 600 nm longpass (FELH0600) and an 850 nm shortpass (FESH0850) filter from Thorlabs at the front of the system.
While the RGB camera is not central to the scope of this study, it is included in the dual-head setup and consists of a Sony IMX420 sensor equipped with a 25mm lens.

\subsection{Spatial and Spectral Resolution Datacube}
The system outputs 3D hyperspectral datacubes $I(x,y,\lambda)$, where the intensity $I$ is a function of the spatial coordinates $(x,y)$ and wavelenght $\lambda$, with spatial dimensions of $312\times \SI{312}{pixels}$ and either 145 or 236 spectral channels depending on whether the reconstruction algorithm used is based on CNN or EM as presented in the following subsection. 

At the expected float altitude of 26 km the system has a ground sampling distance (GSD) of approximately 3.4 $m$ per pixel.

The focal length and spatial resolution desired was motivated to allow for tree type classification in future flight campaigns where the crown spread of e.g. adult spruce specimens typically ranges from 7 to 9 $m^2$, \citep{mortonarboretum_norwayspruce}, although tree clustering is anticipated to increase the area over which specimens vary. Similarly the spectral resolution for tree type classification would need to be $\leq$10 nm to ensure that relevant absorption features are distinguishable, e.g. the chlorophyll a and b peaks  at $662$ nm and $642$ nm respectively \citep{tree_cluster}.

\subsection{CTIS Image reconstruction}
\textit{{Portions of this section are adapted from} \citep{PETERS2025126017}}.
\vspace{0.5cm}

\begin{figure}[H] 
    \centering
    \includegraphics[width=0.9\textwidth]{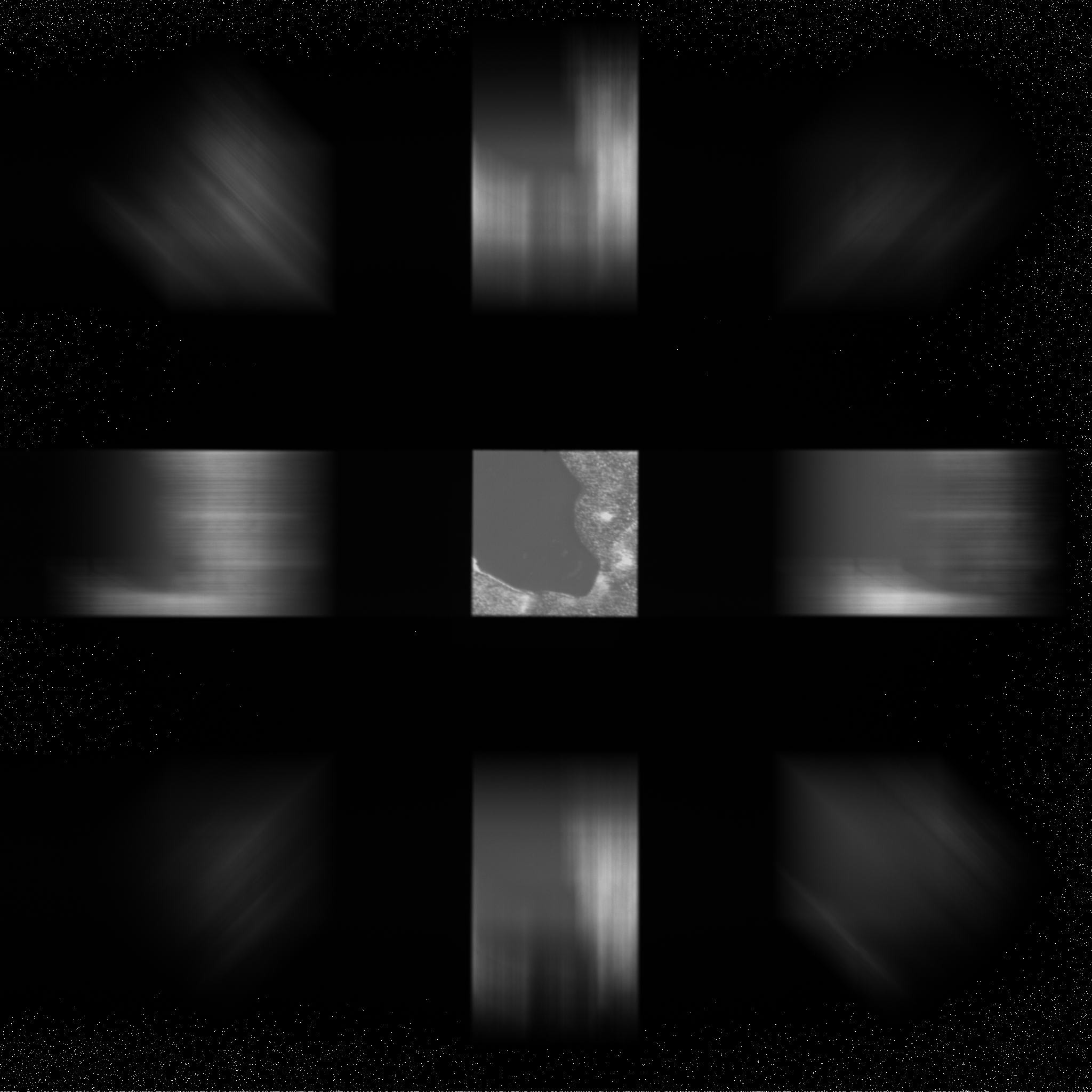}
    \caption{Example CTIS image from the flight. This is image '12255' at approximately 12-13km heigth, resulting in roughly 1.7m/pixel GSD}
    \label{fig:CTISExample}
\end{figure}

The CTIS captures the spatial and spectral information of the data cube simultaneously by utilizing a 2D diffraction grating to disperse the light into a $3\times3$ diffraction pattern of a central zeroth order and 8 surrounding first orders as shown in Fig~\ref{fig:CTISExample}.

From this diffraction image, a tomographic reconstruction is required to obtain the hyperspectral datacube \citep{descour_computed-tomography_1995,Th,Okamoto:91}
The relation between the CTIS diffraction image and the datacube is described by the linear imaging equation that maps the {\it datacube} in vectorized form, $\bm{f}$, 
onto a CTIS diffraction image in vectorized form $\bm{g}$ using the system matrix $\bm{H}$~\citep{descour_computed-tomography_1995}: 
\begin{equation}
    \bm{g}=\bm{H} \bm{f} + \bm{n}
\end{equation}\label{eq:ctis}
with linear noise $\bm{n}$.
In our case $\bm{f}$ is a vector of $312\times312\times(145 $ or $ 236)$ voxels and $\bm{g}$ is a vector of $1910\times{1910}$ pixels. 
The system matrix $\bm{H}$ incorporates the diffraction sensitivity (lens transmission, sensor response, and diffraction efficiency of the diffractive optical element), illumination, and vignetting.
For non-trivial system dimensions, the CTIS system is an underdetermined linear equation with no exact solutions. 
Therefore, either iterative reconstruction algorithms such as the EM algorithm~\citep{descour_computed-tomography_1995, white_accelerating_2020} or CNNs~\citep{huang_application_2022,ahlebaek_hybrid_2023} are utilized to approximate the inverse of $\bm{H}$ and reconstruct datacubes from CTIS images. Our description and implementation of the EM algorithm follows~\citep{huang_application_2022} and consists of an expectation step, where we compute $\hat{\bm{g}} = \bm{H} \hat{\bm{f}}^{(k)}$, and a maximization step, where we update $\hat{\bm{f}}^{(k)}$:
\begin{equation}
	\hat{\bm{f}}^{(k+1)}	 = \frac{\hat{\bm{f}}^{(k)}}{\sum_{i=1}^{q^2} H_{ij}}\odot \left( \bm{H}^T \frac{\bm{g}}{\bm{H}\hat{\bm{f}}^{(k)}}\right)
\end{equation}\label{eq:EM}
Here, $k$ is the iteration index, $\hat{\bm{f}}^{(k)}$ is the current estimate, $\sum_{i=1}^{q^2} H_{ij}$ is the sum of rows in $\bm{H}$, $\bm{H}^T$ is the transpose of $\bm{H}$, and $\odot$ denotes element-wise multiplication. This equation combines the expectation and maximization steps. We initialize with $\hat{\bm{f}}^{(0)} = \bm{H}^T\bm{g}$ and perform 20 iterations, as typically 10-30 are needed~\citep{white_accelerating_2020}. Both the $\bm{H}$ construction and $\bm{f}$ reconstruction are implemented in \texttt{MATLAB} using sparse matrix manipulations. As detailed in section~\ref{sec:CTISSystem}, the optical system of the CTIS cameras is the same -- with the sole exception of the outermost lens -- as the system outlined in~\citep{ahlebaek_hybrid_2023}, which also applies to the construction of the $\bm{H}$ matrix. A series of experiments were carried out to precisely replicate the inclusion of spatial and spectral correction terms (such as point spread function, illumination, quantum efficiency, etc.) as elaborated in the supplementary material of~\citep{ahlebaek_hybrid_2023}.

CNNs have previously been used for classification directly on CTIS images~\citep{douarre_ctis-net_2021} and to reconstruct datacubes from CTIS images~\citep{huang_application_2022,ahlebaek_hybrid_2023,mel2022joint,Wu,Zimmermann}, and as an extension to that, we proposed an autodecoder network architecture in~\citep{ PETERS2025126017}. This physics-guided convolutional autodecoder consists of the UNet presented in~\citep{huang_application_2022} as the decoder, followed by an encoder that utilizes physical knowledge of the imaging system contained in the system matrix $\bm{H}$. It utilizes the linear imaging equation (Eq.~\ref{eq:ctis}) to obtain an estimated CTIS image $\hat{\bm{g}} = \bm{\hat{H}}\bm{f}$ (since the exact system matrix $\bm{H}$ is unknown), and an additional UNet structure is employed that refines the estimated CTIS image.

\section{Imaging System Integration}
\label{Imaging System Integration}
In this section we detail some of the mechanical and thermal requirements and designs in order for the imaging system to function on the BEXUS HAB platform and in stratospheric conditions 

The trajectory of the HAB flight is shown in Figure~\ref{fig:Traj} and passes over both land and water bodies. The flight was planned around solar noon to have uniform lighting conditions for the test flight. 
During flight, the ballon reaches 28 km altitude as shown in Figure~\ref{fig:Altitude}. At this height temperatures of $<$ -50 C$^\circ$ as well as a pressure as low as 0,05 atm is reached in which the imaging system must remain operational.

\subsection{Mechanical}
The robustness and stability of the mechanical gondola and payload structure are crucial for HAB flights. A structural failure could result in experiment failure or even the loss of equipment mid-flight. Although most of the flight is smooth, three critical events can subject the structure to forces up to 20 $g$: balloon cutdown, parachute engagement, and gondola touchdown.

\begin{figure}[H]
\centering
\includegraphics[width=0.8\linewidth]{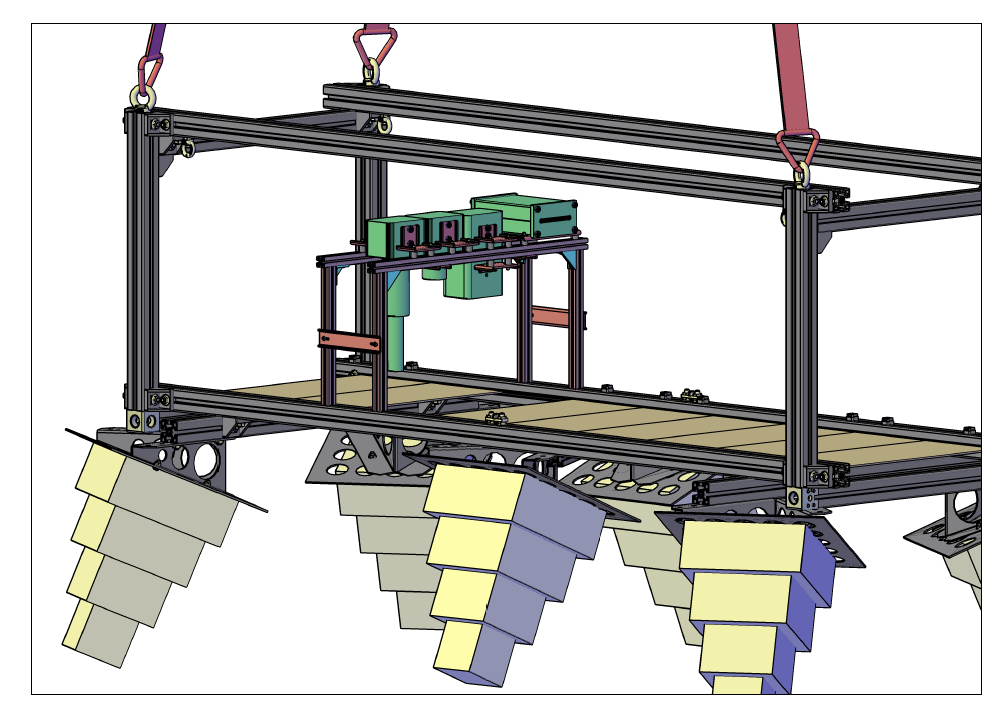}
\caption{CAD model of the experiment fully integrated on the BEXUS gondola.}
\label{fig:IntegratedStructure}
\end{figure}

An overview of the BEXUS gondola with the imaging system for the HEIMDAL mission is shown in Figure~\ref{fig:IntegratedStructure}. The four green modules are, from left to right, the CTIS camera, an RGB Camera, the on-board computer (OBC) and the electronics box.

The Gondola is built by 45x45 mm Rexroth aluminum profiles, assembled by various angle gussets in the corner interfaces, with an effective volume of 2x0.6x0.6 (LxWxH) m$^3$.

\begin{figure}[H]
\centering
\includegraphics[width=0.8\textwidth]{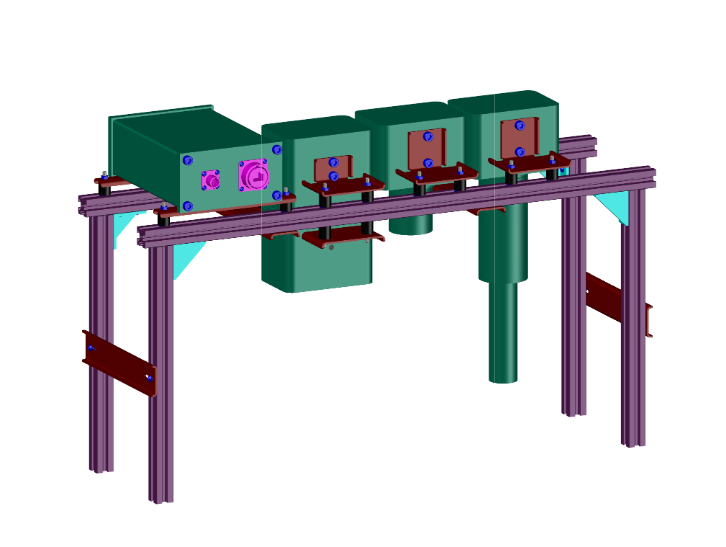}
\caption{CAD model of the experiment. Showing the four modules, the experiment frames, braces and brackets.}
\label{fig:HEIMDAL_Mecha}
\end{figure}

The mechanical structure for the imaging system was designed using six Rexroth 20×20 mm profiles (See Figure. \ref{fig:HEIMDAL_Mecha}): four vertical 470 mm legs to elevate the four experiment modules from the gondola frame and two horizontal 600 mm legs connecting the four 470 mm legs. These six profiles were secured using corner gussets and additional custom braces that connect the leg frames for stability.

These custom braces were initially designed and manufactured as 2 mm aluminum bent sheet metal designs, but following a shock test, these were deemed too soft, and a new set of braces were manufactured out of steel.

Rubber buffers were used for shock absorption and electrical insulation to mount the experiment modules. 

The four modules were all fastened by T-nuts to the rails, through the rubber buffers onto respective component brackets. Additionally, Loctite was used to ensure nothing came loose during launch.
The electronics box was wrapped in Multi-Layer Insulation (MLI) to protect the SSDs housed inside, as deemed necessary by the the first thermal vacuum chamber test in the Mars Lab at Aarhus University.
The final integrated structure can be seen in Figure~\ref{fig:AssemblyLaunchCampaign},
which also shows that the frame beneath the electronics box is wrapped in brown insulation tape. This precaution was necessary because MLI is electrically conductive, and direct contact with the frame could have caused a short circuit.

The final total weight of the payload is 7.5 kg, with the dimensions: 0.60 × 0.25 × 0.47  m$^3$.

\subsection{Software, Camera control, Image storage}
\label{sec:camcontrol}
The flight software for the mission consisted of two separate systems: For the OBC and software for the groundstation control room. Furthermore, the OBC software was split into two parts: Image storage and reconstruction, and camera control. The communication between the OBC and groundstation was enabled by an ethernet interface, called E-link provided by the launch provider, Swedish Space Cooperation (SSC).

Images were captured and stored every second on three separate SSD drives during flight, with each image assigned an incrementing suffix number such as image 12255 shown in Fig~\ref{fig:CTISExample}. To validate on board image reconstruction, while at the same time limiting power consumption we reconstructed data cubes from CTIS images on-board for every 10th image, yielding around 2000 cubes. A fraction of the images were also downlinked during flight through the E-LINK at a rate of 1 image per 90 seconds. 

\begin{figure}[H] 
    \centering
    \includegraphics[width=0.9\textwidth]{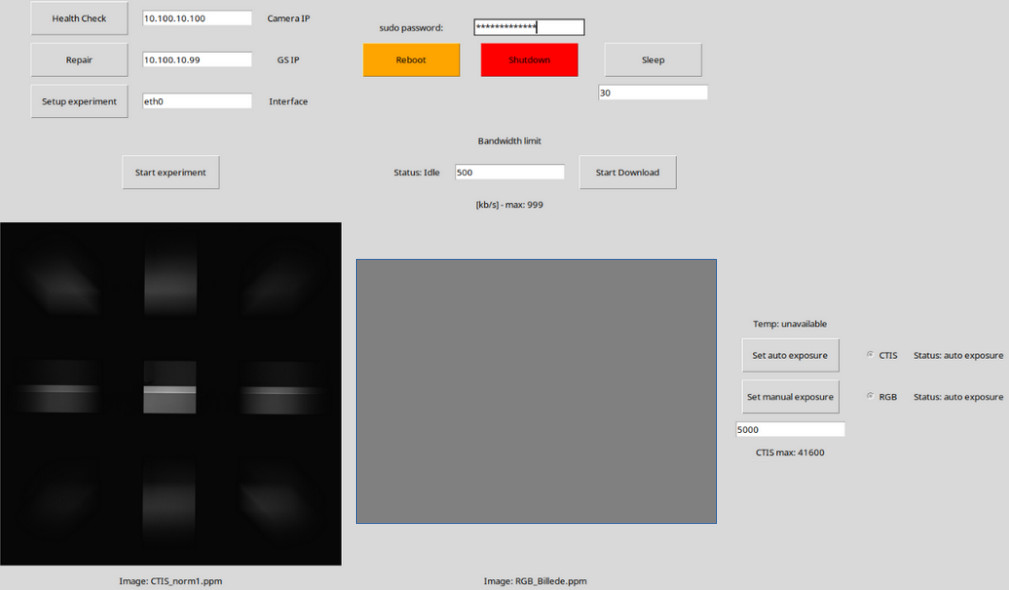}
    \caption{A snapshot of the GUI used during flight operations.}
    \label{fig:gui}
\end{figure}

A Graphical User Interface (GUI) was developed to monitor and control the imaging system during flight. A snapshot of the GUI is shown in Figure~\ref{fig:gui}. The GUI allowed the operations team to change between manual- and auto-exposure, control the bandwidth with which the images were transmitted to the ground station, and to monitor the image quality visually from the downlinked images  

\subsection{Power system}
The power consumption of the camera system, incl. OBC was estimated to be 180 Wh, budgeting for 1 hour in idle mode prior to launch and five active hours during flight.
Power was provided by 2 parallel connected battery packs with 8x SAFT LSH20 (3.6V) batteries configured in series and delivering a nominal 28.8 V with a capacity of 13000 mAh corresponding to 750 Wh at $20^\circ$C.
The expected nominal temperature inside the battery compartment during flight is $-20^\circ$C which reduces the effective capacity of one battery pack to roughly 173 Wh. We therefore use two battery packs to ensure sufficient power for the entire flight.

Moreover, the battery pack delivered an unstable 28.8 V, while the OBC demands a stable voltage in the range 20-28.8V, with tests done at 24 V. 
Voltage regulation was achieved with a Power Distribution Unit (PDU), 
consisting of a TDK-Lambda i7C Buck-Boost converter, upon which the PDU was designed in KiCad and produced by ICAPE Denmark A/S, lowering the operational voltage to 24 V. 

\begin{figure}[H]
    \centering
    \includegraphics[width=0.5\linewidth]{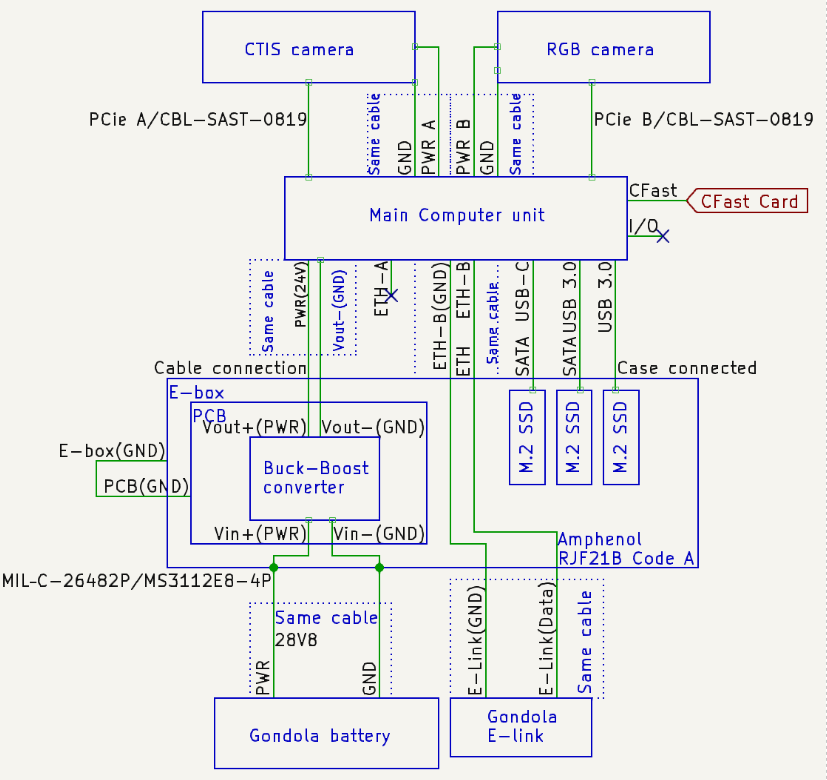}
    \caption{Full power system design including all major components, connections, and power distribution and data pathways.}
    \label{fig: Power system}
\end{figure}

The complete electronics design is shown in Figure~\ref{fig: Power system}.

\subsection{Pre-flight tests}
The most significant concern regarding the system performance was the thermal design surrounding the OBC and the cameras on-board computing units, particularly whether the experiment would overheat in the low-pressure environment of the stratosphere due to limited convective cooling. The high power consumption of the prototype camera, combined with low heat dissipation, raised concerns about potential thermal bottlenecks.

\begin{figure}[H]
    \centering
    \begin{subfigure}[b]{0.45\textwidth}
    \includegraphics[width=\textwidth]{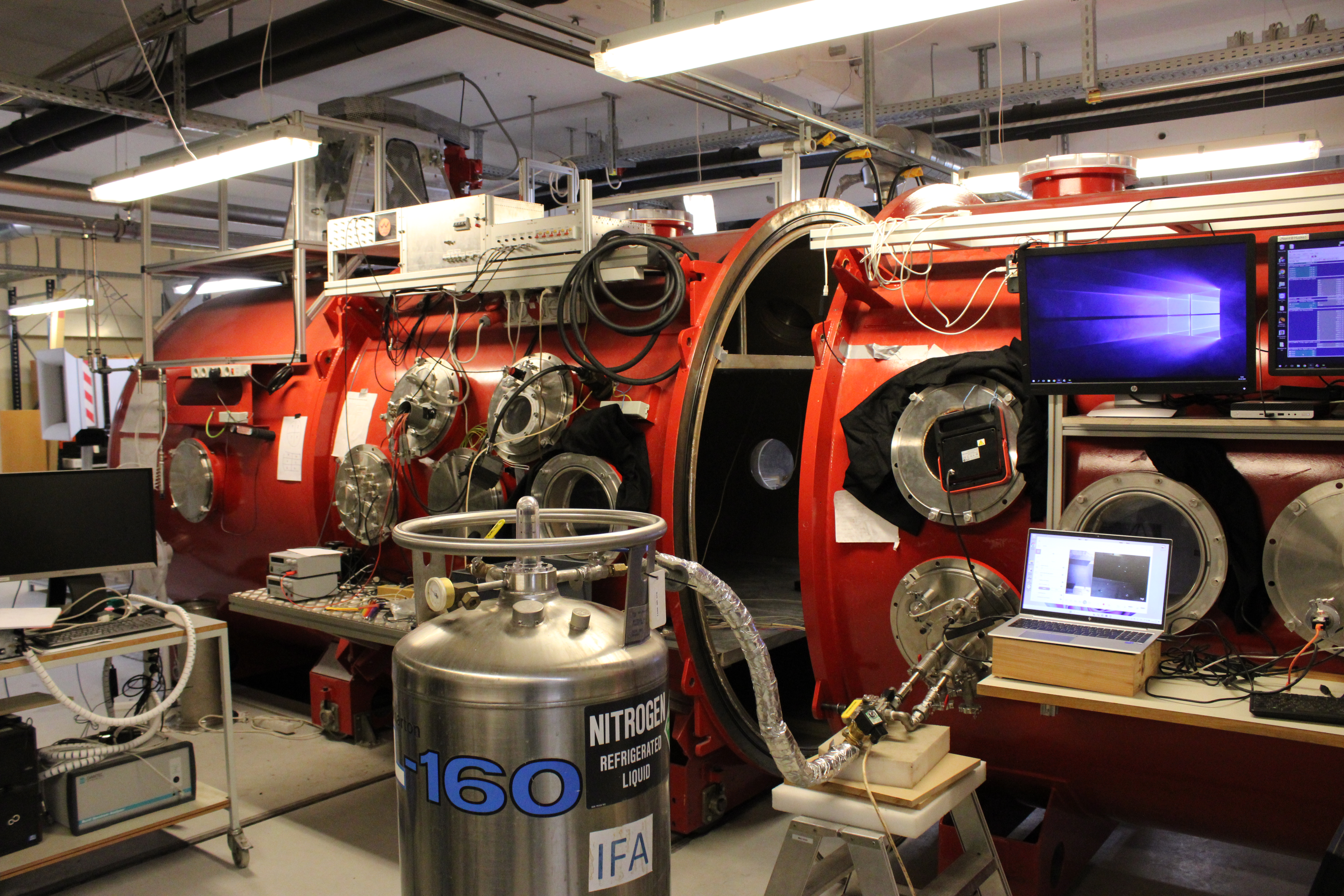}
    \caption{}
    \end{subfigure}
    \hfill
    \begin{subfigure}[b]{0.535\textwidth}
    \includegraphics[width=\textwidth]{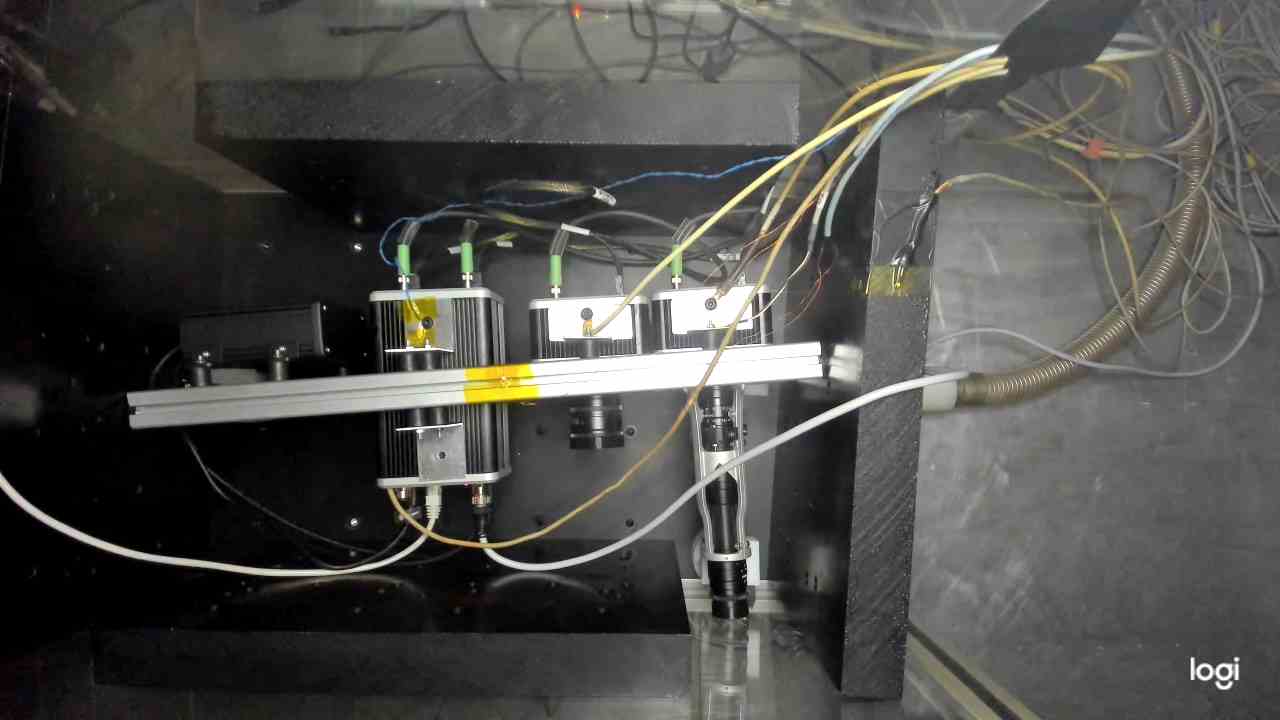}
    \caption{}
    \end{subfigure}
    \caption{Panel (a) shows the exterior of the Mars Simulation Laboratory vacuum chamber, while (b) shows the experiment inside the vacuum chamber as the first test is ongoing.}
    \label{fig:MarsLab}
\end{figure}

To address these concerns, the experiment was tested in the Mars Simulation Laboratory at Aarhus University. This state-of-the-art vacuum chamber is designed to replicate Martian environmental conditions and is equipped with a cooling plate that circulates liquid nitrogen to achieve extremely low temperatures. Figure~\ref{fig:MarsLab}, panel (a) shows the exterior of the Mars Laboratory vacuum chamber, while panel (b) presents the experiment, while the thermal test is in progress.
The test protocol was to subject the experiment to a 5 mbar pressure environment while varying the environmental temperature between -20°C and -80°C for over 5 hours. These test conditions were designed to simulate conservative worst-case scenarios for pressure, temperature, and duration.
The first test revealed that our SSDs would not withstand the harsh environment, prompting us to wrap the electronics box in multilayer insulation. This effectively reduces the radiated heat, helping to maintain an operational temperature of the electronics for a longer duration.
A follow-up test was performed after the insulation was added, which confirmed that the experiment remained fully operational throughout the entire thermal vacuum test.

\section{Flight and Operations} \label{sec:operations}
The system was powered on several hours before launch to conduct communications and systems checks, resulting in approximately 9,500 images being captured while the gondola was still on the ground. During this period, the gondola was powered by a cable to prevent battery consumption. As a result, the images presented in the results section have numbers ranging from 10,000 to 12,000, but these are still from the first 500-2500 seconds of the flight.

\begin{figure}[H] 
    \centering
    \includegraphics[width=\textwidth]{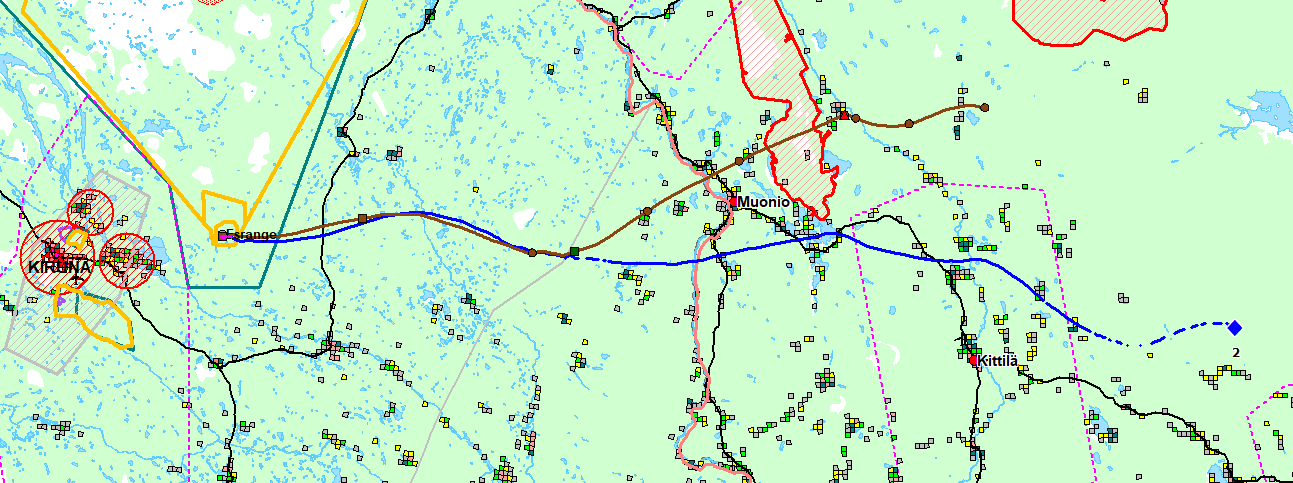}
    \caption{The trajectory is plotted on a map displaying cities, flight zones, and the border between Sweden and Finland. The dashed pink lines represent "No-Landing" zones, while the yellow area indicates a restricted zone. The red regions are designated nature reserves. The blue line represents the actual flight path, and the brown line depicts the predicted trajectory.}
    \label{fig:Traj}
\end{figure}

The balloon was launched at 07:52 on October 2nd from the Balloon Platform at Esrange. The predicted (brown) and actual (blue) flight paths are shown in Figure~\ref{fig:Traj}.
Initially, the paths align closely, but after about 1.5 hours, they begin to diverge. The flight, which started over Sweden, concluded in a forested area of Finnish Lapland. Although the planned duration was approximately three hours, as the balloon deviated from the predicted trajectroy it entered a "No-Landing" zone, resulting in an extended total flight time of around 5.5 hours.

\begin{figure}[H] 
    \centering
    \includegraphics[width=\textwidth]{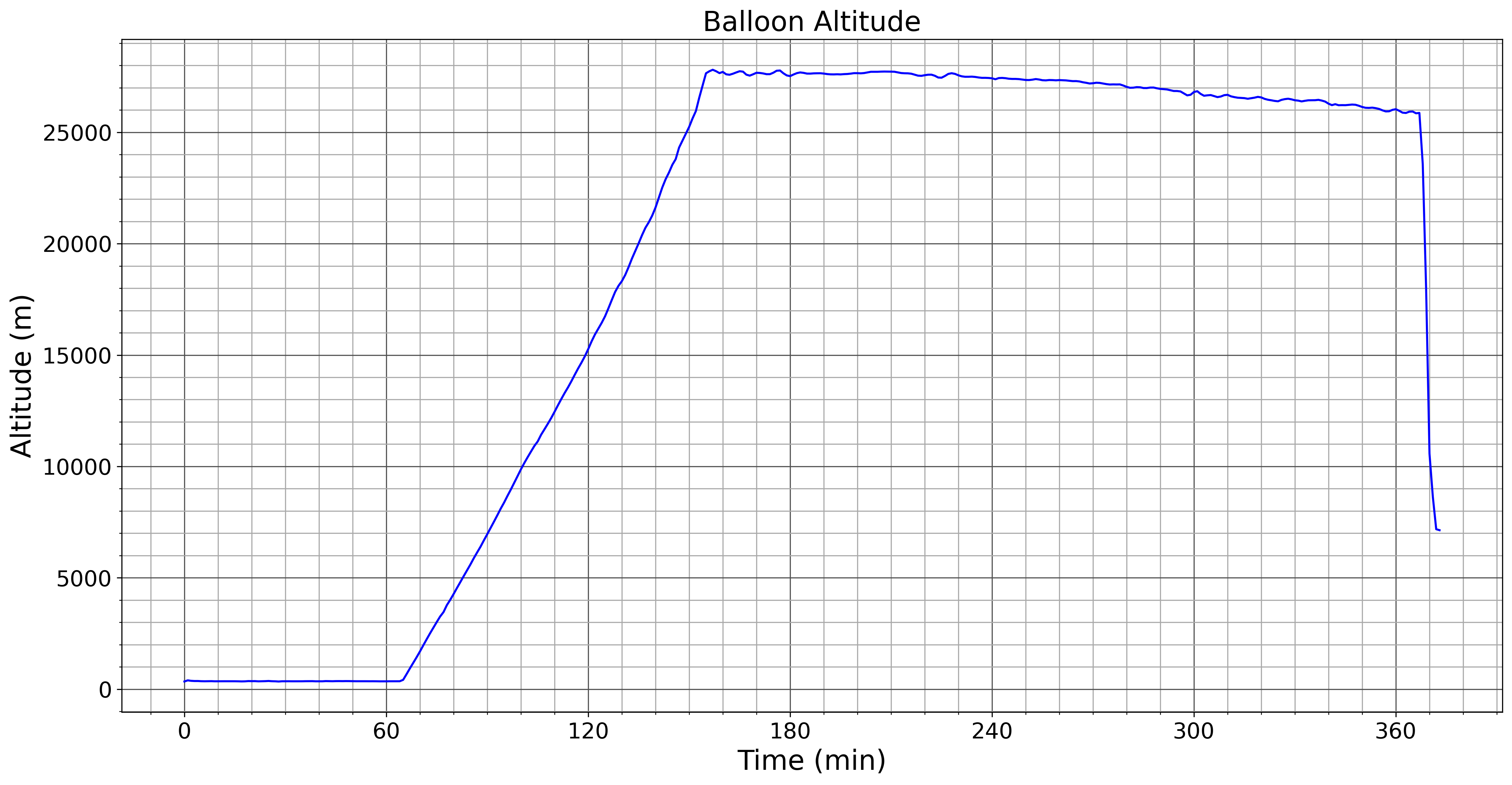}
    \caption{Plot showing the altitude (m) vs. time (min) for the flight. Altitude is shown as height over mean sea level.}
    \label{fig:Altitude}
\end{figure}

The altitude profile is shown in Figure~\ref{fig:Altitude} and presents a very linear ascent profile of roughly 4.8m/s as expected, but the balloon reached a higher maximal height than expected at ca. 28 km.
This could be the reason as to why the predicted trajectory deviated from the actual trajectory. 

\section{Results}
\label{sec:results}
Both cameras  captured images during the flight and all images were safely stored on the SSDs that survived the descend and landing of the gondola. However a number of issues also arose during the flight. 

It should be noted that all images presented in the Results section and Appendix A were acquired during the ascent phase of the balloon flight. Consequently, the ground sampling distance (GSD) varies slightly between images due to changes in altitude. From the shown images, the highest spatial resolution is achieved in image 11367, with an estimated GSD of approximately 1.5 m/pixel, while the lowest resolution is observed in image 12277, with a GSD of approximately 2.1 m/pixel. These GSD values are theoretically derived based on the sensor’s pixel size, the focal length of the optical system, and the corresponding acquisition altitude.

\subsection{RGB Camera}
\begin{figure}[H] 
    \centering
    \includegraphics[width=0.9\textwidth]{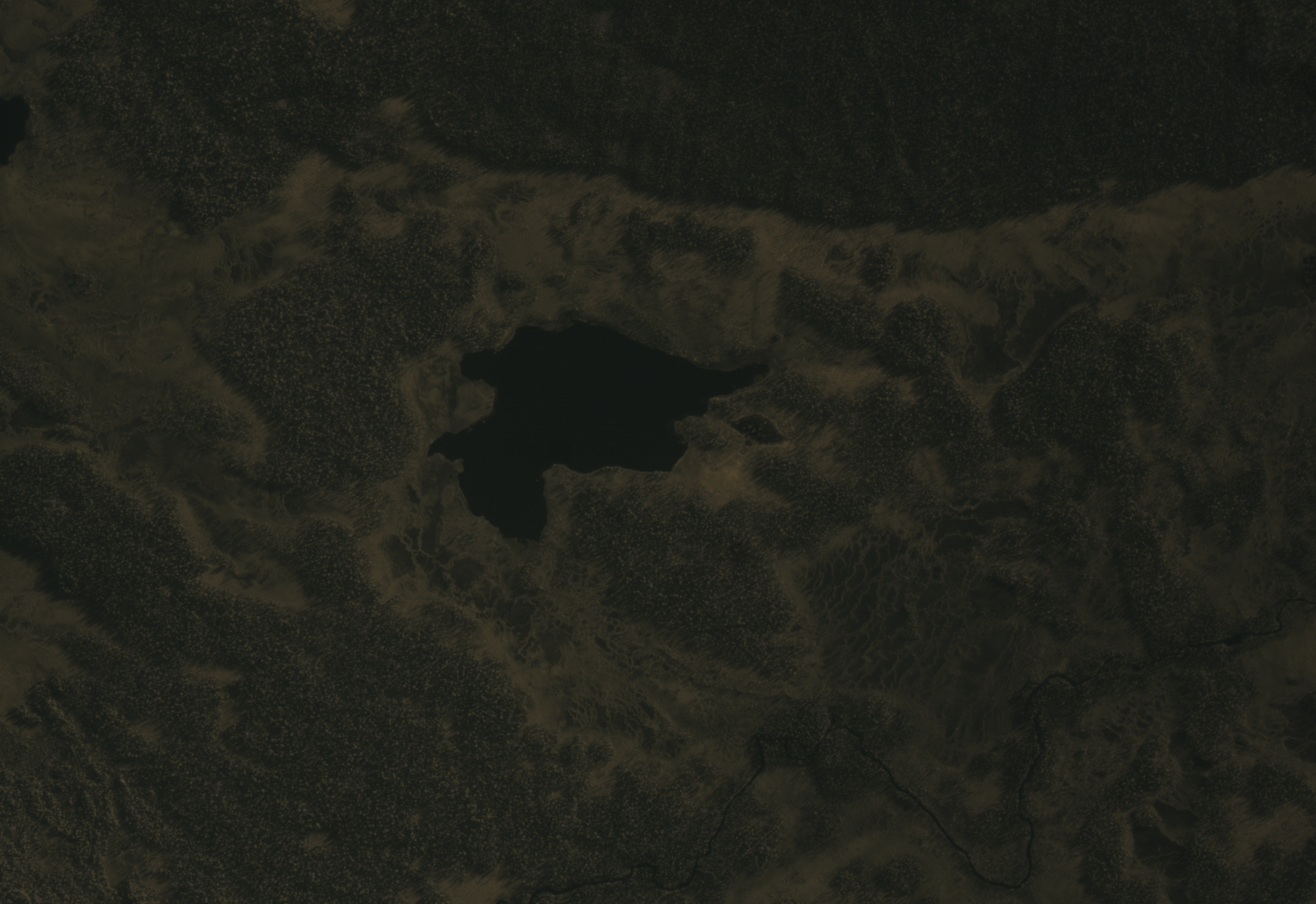}
    \caption{Example of an acquired RGB image, showcasing the low exposure. This is image '11782' at approximately 10-11km altitude with roughly 1.9 m/pixel GSD.}
    \label{fig:RGBExample}
\end{figure}

The auto-exposure function did not operate correctly on the RGB camera, leading to consistently underexposed images which can be seen in an example RGB image shown in Figure~\ref{fig:RGBExample}.
Although the camera captured images at 1 FPS, the transmission limit of 500kbps resulted in an image delay at the ground station of approximately 90 seconds, which significantly hindered real-time assessment and manual exposure adjustments during flight.

A second issue encountered was that some of the RGB images were completely black, likely originating from the trigger signal synchronization between the RGB and CTIS cameras: The cameras were configured in a master-slave setup, and a possible misalignment in their frame rates may have resulted in missed or improperly captured exposures.

Lastly, towards the latter part of the flight, condensation became visible in the captured images, causing a hazy appearance and introducing a distinct artifact in the top-right corner of the frames.

\subsection{CTIS Camera}
The CTIS camera operated smoothly throughout the flight, with no major technical issues. However, similarly to the RGB camera, condensation became apparent during the later stages of the flight, heavily impacting the image quality.
An example CTIS image from the early stages of the flight is shown in Figure. \ref{fig:CTISExample}.

Both the RGB camera and the CTIS camera captured images at 1 FPS, with exposure times between 4–10ms. These short exposure times effectively mitigate any potential motion blur caused by gondola rotation or swinging.

\begin{figure}[H] 
    \centering
    \includegraphics[width=0.9\textwidth]{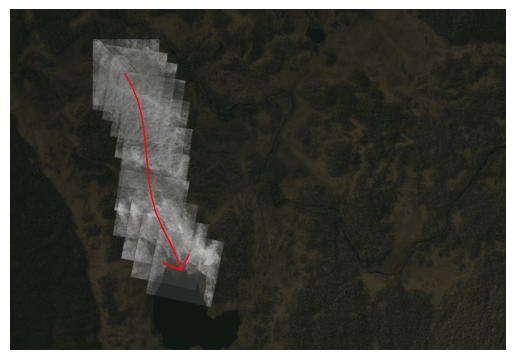}
    \caption{Shows the 0th order of CTIS images 11770-11800 overlaid on a corresponding RGB image during flight. The red arrow illustrates the motion of the balloon.}
    \label{fig:path}
\end{figure}

In Figure~\ref{fig:path}, the 0th order of CTIS images 11770–11800 is overlaid on a corresponding RGB image captured by the wider Field of View (FOV) RGB camera. This overlay highlights the motion between each frame caused by the gondola's rotation and pendular oscillations during flight, demonstrating that the snapshot HSI technique is the most suitable method for hyperspectral imaging acquisition under these conditions.
This overlaying is done by the use of an autocorrelation algorithm which is a feature-based registration approach using Oriented FAST and Rotated BRIEF (ORB) keypoints. The method consists of four primary steps: feature detection and matching (BruteForceMatcher), homography estimation, transformation parameter extraction, and image warping.

\begin{figure}[H]
    \centering
    \begin{subfigure}{0.32\textwidth}
        \centering
        \includegraphics[width=\textwidth]{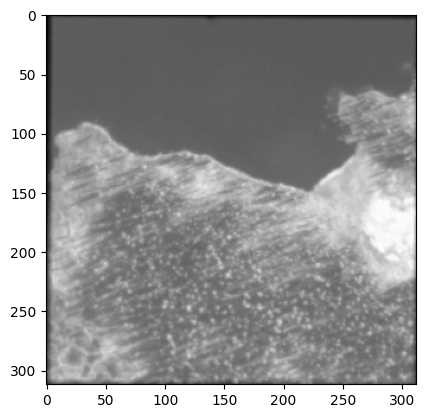}
        \caption{Shows the 0th order of a CTIS image containing both water and land used for training is.}
        \label{fig:PLSDACTISNoMask}
    \end{subfigure}
    \hfill
    \begin{subfigure}{0.32\textwidth}
        \centering
        \includegraphics[width=\textwidth]{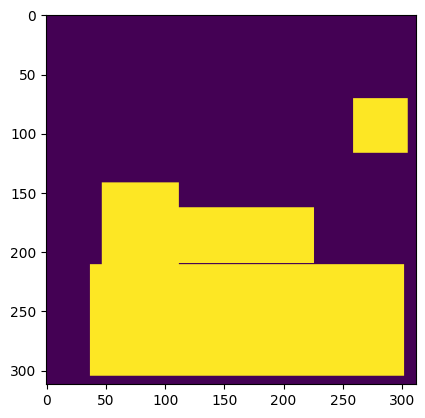}
        \caption{Shows the mask from (a) including land. \vspace{10pt}}
        \label{fig:PLSDATreeMask}
    \end{subfigure}
    \hfill
    \begin{subfigure}{0.32\textwidth}
        \centering
        \includegraphics[width=\textwidth]{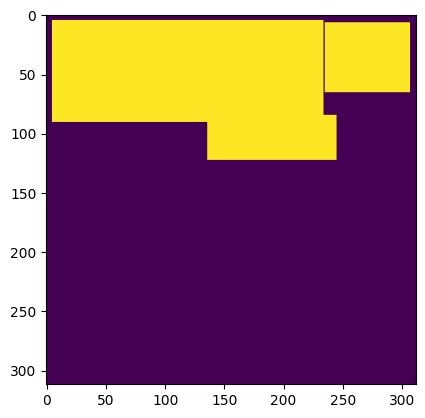}
        \caption{Shows the mask from (a) including water.\vspace{10pt}}
        \label{fig:PLSDAWaterMask}
    \end{subfigure}
    
    \caption{Example of the manually created masks including land and water for the training of the PLS-DA models.}
    \label{fig:PLSDATrainMask}
\end{figure}

As outlined in Sec.~\ref{sec:CTISSystem}, the raw CTIS images are 2D diffraction images, and require image reconstruction to obtain the datacube. To do this both a CNN-based and an EM-based reconstruction algorithm is used. To provide a simple proof-of-principle analysis 23 datacubes with areas of both water and land were manually selected and masked out. One of these 23 images and the corresponding masks are shown in Figure~\ref{fig:PLSDATrainMask}.

The spectra from these regions were averaged over 5×5 neighboring pixels to reduce noise, and Standard Normal Variate (SNV) normalization was applied for standardization. The resulting (training) dataset consists of 213,206 spectra across 23 images, with 72\% representing tree-covered areas and 28\% representing water, 
reflecting the prevalence of land regions in the observed area makes this imbalance expected.

Using the normalized spectra a Partial Least Squares Discriminant Analysis (PLS-DA) was performed to qualitatively assess the ability to differentiate between land and water in the datacubes.

\begin{figure}[H]
    \centering
    \begin{subfigure}[b]{0.45\textwidth}
    \includegraphics[width=\textwidth]{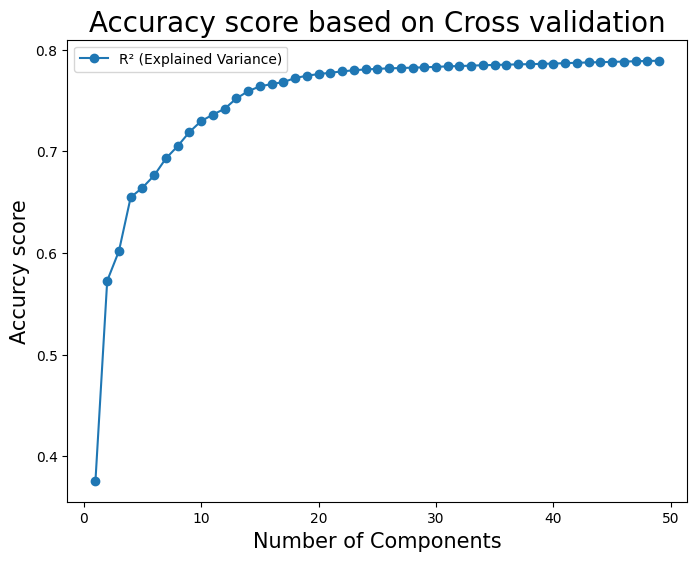}
    \caption{}
    \end{subfigure}
    \hfill
    \begin{subfigure}[b]{0.46\textwidth}
    \includegraphics[width=\textwidth]{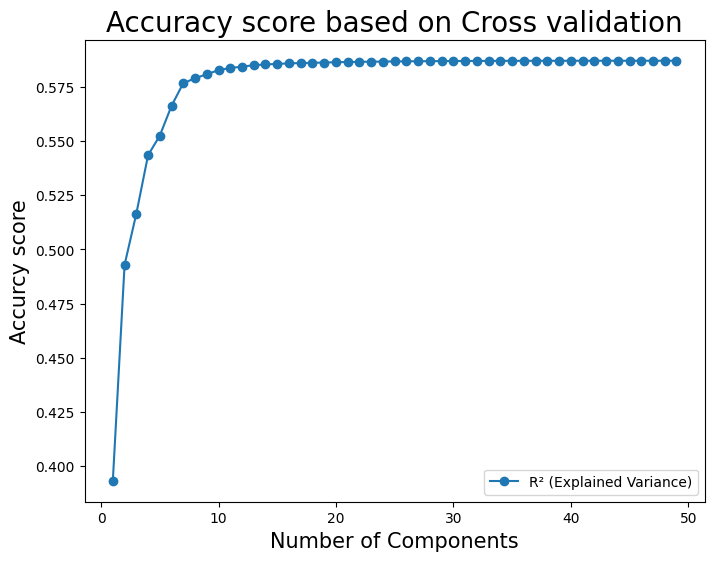}
    \caption{}
    \end{subfigure}
    \caption{Comparison of accuracy scores (R-squared values) for PLS-DA models using CNN and EM reconstructions with an increasing number of components. The CNN-based model (a) converges around 0.8, while the EM-based model (b) reaches a lower final accuracy of approximately 0.58. Both models show diminishing accuracy gains beyond a certain number of components — 25 for CNN and 15 for EM — indicating a point of diminishing returns in model complexity.}
    \label{fig:AccuracyScore}
\end{figure}

Two PLS-DA models are trained using reconstructed datacubes from each of the two reconstruction methods, and a Cross validation accuracy score for each model is shown in Figure~\ref{fig:AccuracyScore}. The accuracy scores indicate comparable performance between the models; with the CNN-based converging to and accuracy score of 0.8 with ~25 components, whereas the EM-based model converges to 0.58 with only 15 components. This suggests that the CNN-based model will outperform the EM-based model.

\begin{figure}[H]
  \centering
  \begin{subfigure}[b]{0.45\textwidth}
    \includegraphics[width=\textwidth]{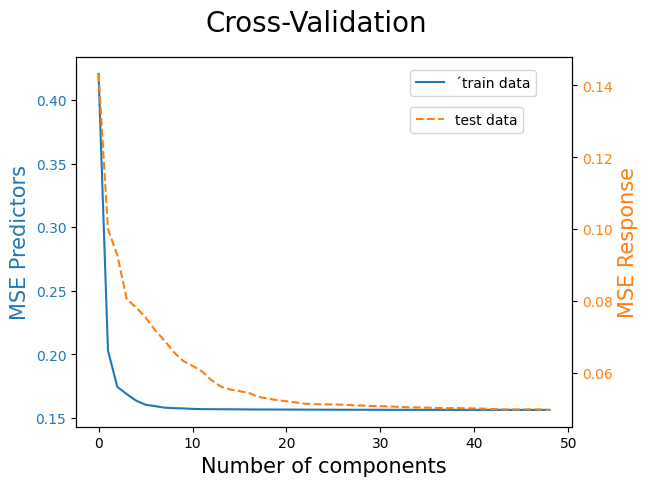}
    \caption{}
  \end{subfigure}
  \hfill
  \begin{subfigure}[b]{0.455\textwidth}
    \includegraphics[width=\textwidth]{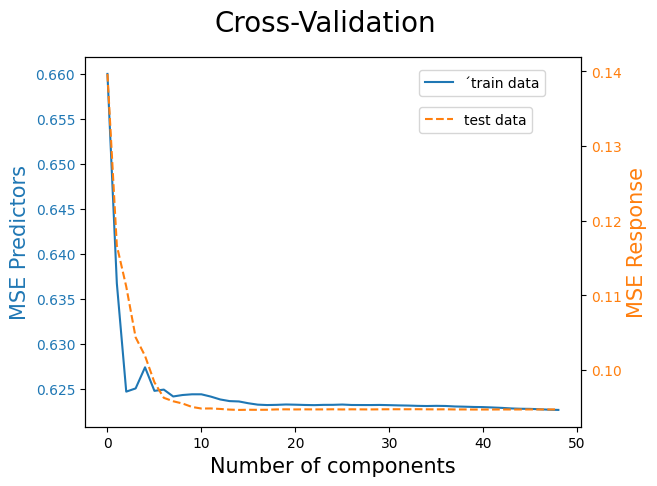}
    \caption{}
  \end{subfigure}
  \caption{Comparison of Mean Squared Error (MSE) of the two as a function of number of components. (a) shows the MSE for training and test data for the CNN-based model, while (b) shows the MSE for the EM-based model.}
  \label{fig:MSEPlots}
\end{figure}

In addition to accuracy scores, the performance of the two models is further evaluated using the Mean Squared Error (MSE) for both training and test data. Figure~\ref{fig:MSEPlots} presents the MSE curves for the CNN- and EM-based models as a function of the number of PLS-DA components. These plots provide insight into the models' generalization ability, where a lower test MSE indicates better predictive performance. Similarly, to the Accuracy scores, this indicates that the CNN based model outperforms the EM-based model.

Beyond accuracy and Mean Squared Error (MSE), another key metric for evaluating the PLS-DA models is the Variable Importance in Projection (VIP) score. The VIP score provides insight into the contribution of each spectral band to the model's predictions, highlighting the most influential wavelengths for distinguishing between land and water.

\begin{figure}[H]
  \centering
  \begin{subfigure}[b]{0.45\textwidth}
    \includegraphics[width=\textwidth]{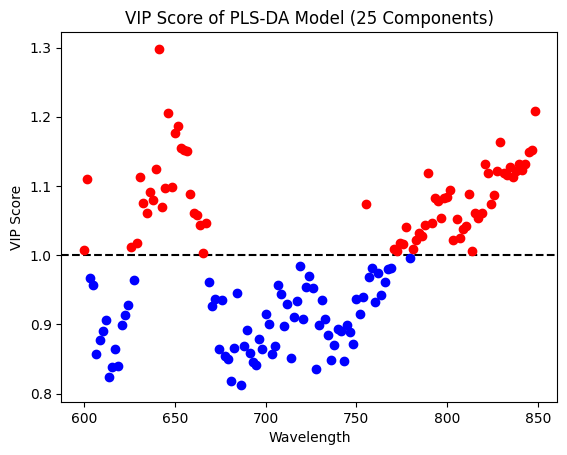}
    \caption{}
  \end{subfigure}
  \hfill
  \begin{subfigure}[b]{0.45\textwidth}
    \includegraphics[width=\textwidth]{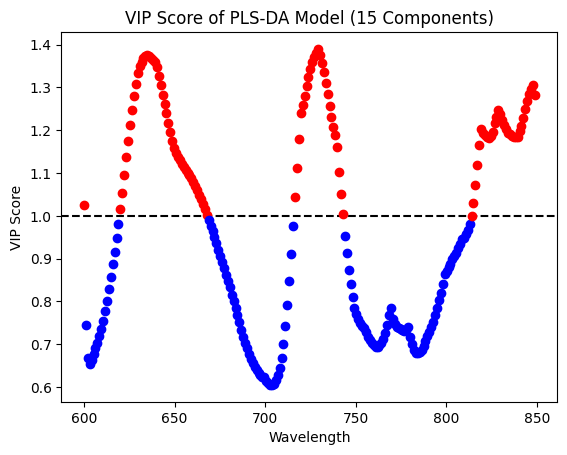}
    \caption{}
  \end{subfigure}
  \caption{Variable Importance in Projection (VIP) scores from PLS-DA traind on CNN-based reconstructions (a), and EM-based reconstructions (b).}
  \label{fig:VIPScores}
\end{figure}

Figure~\ref{fig:VIPScores} presents the VIP scores for the CNN- and EM-based models, illustrating how the importance of different spectral bands varies between the two reconstruction approaches. This analysis helps assess not only the models' performance but also their interpretability in terms of spectral feature selection.

The VIP scores highlight consistently significant spectral features around 650 nm for both models, while showing some variability in importance across the second half of the spectrum. However, this observation by itself is insufficient to conclusively determine a preference for one reconstruction method over the other.

\begin{figure}[H]
    \centering
    \begin{subfigure}[b]{0.44\textwidth} 
        \centering
        \includegraphics[width=\textwidth]{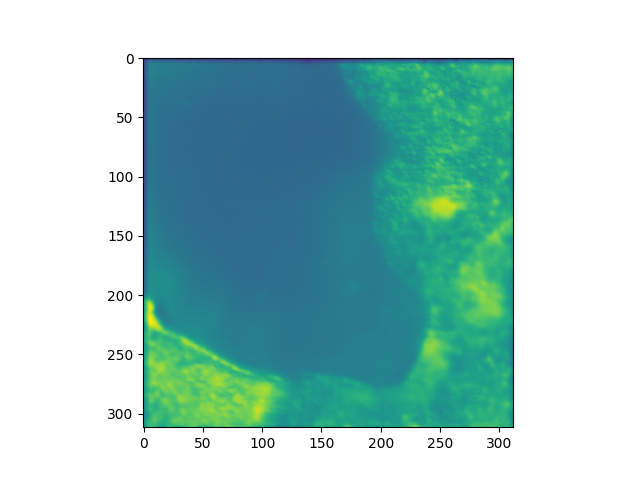}
        \caption{Shows the spectral averaged CNN reconstruction of image 12255}   
        \label{fig:mean and std of net14}
    \end{subfigure}
    \hfill
    \begin{subfigure}[b]{0.55\textwidth}  
        \centering 
        \raisebox{0.35cm}{\includegraphics[width=\textwidth]{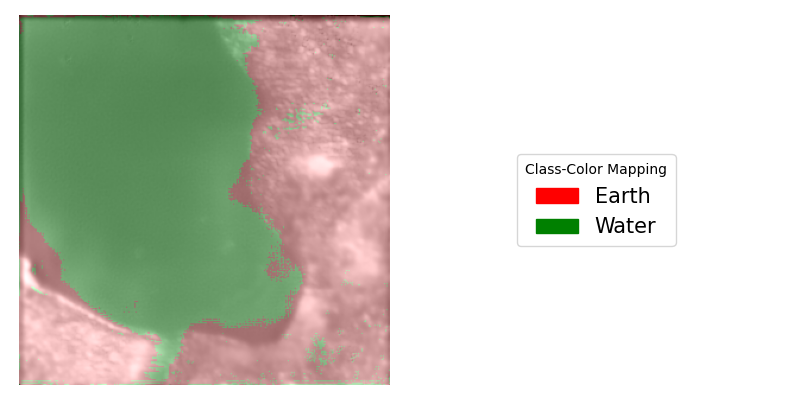}} % Adjust height
        \caption{Shows the CNN-based PLS-DA prediction map using 25 components, as a transparent overlay over (a)}
        \label{fig:mean and std of net24}
    \end{subfigure}
    
    \vskip \baselineskip
    
    \begin{subfigure}[b]{0.44\textwidth}   
        \centering 
        \includegraphics[width=\textwidth]{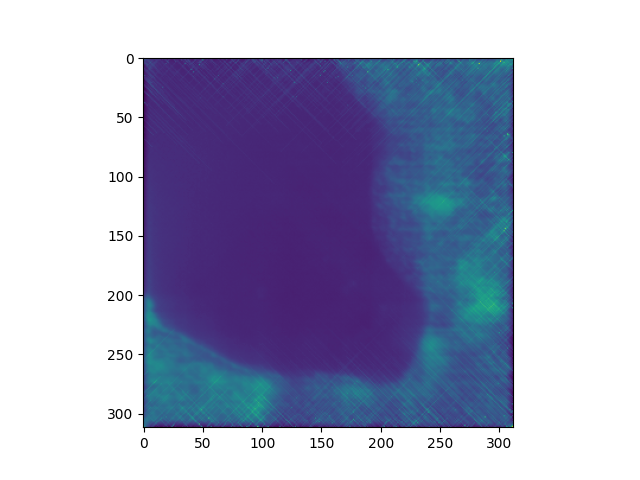}
        \caption{Shows the spectral averaged EM reconstruction of image 12255}    
        \label{fig:mean and std of net34}
    \end{subfigure}
    \hfill
    \begin{subfigure}[b]{0.55\textwidth}   
        \centering 
        \raisebox{0.35cm}{\includegraphics[width=\textwidth]{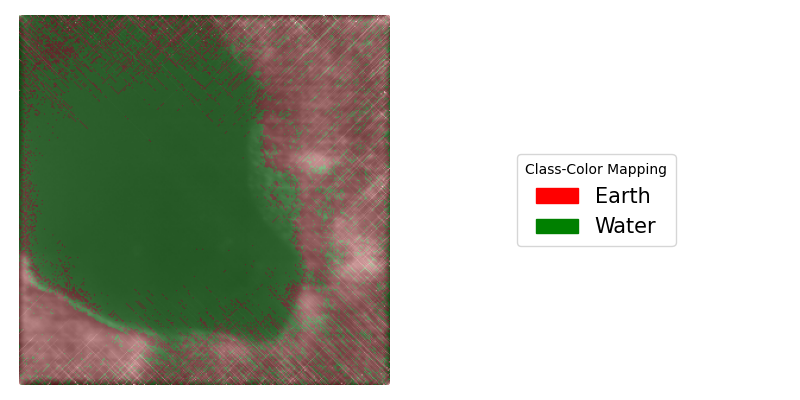}} % Adjust height
        \caption{Shows the EM-based PLS-DA prediction map using 15 components, as a transparent overlay over (c)} 
        \label{fig:mean and std of net44}
    \end{subfigure}
    
    \caption{CNN and EM reconstructions and corresponding PLS-DA prediction}
    \label{fig:mean and std of nets}
\end{figure}

In Figure~\ref{fig:mean and std of nets}, panel (a) presents the CTIS image 12255 reconstruction based on the CNN model, where all spectral bands are averaged to form a 2D image. This image is visualized using the Viridis colormap. The image is not part of the training dataset and was selected as an example containing both land and water. Panel (c) shows the corresponding EM reconstruction of the same CTIS image, with spectral averaging applied in the same manner. Panels (b) and (d) present the corresponding PLS-DA results based on the CNN and EM reconstructions, respectively. Upon inspection, both models successfully identify the presence of both water and land and generally map these regions correctly. However, the EM reconstruction exhibits some noticeable artifacts, such as distinct lines within the datacube.
Additionally, upon closer inspection, it is evident that neither model accurately captures the precise waterline. This suggests that the HSI camera may capture spectral information from the surface just below the water, which resembles the characteristics of the land, potentially leading to misclassifications by the PLS-DA models and making it more difficult to correctly delineate the waterline.

Additional figures, including Figure~\ref{fig:mean and std of nets}, are provided in \ref{sec:AppA} for different CTIS images used as input, offering further analysis and validation.

\begin{figure}[H]
    \centering
    \begin{subfigure}{0.32\textwidth}
        \centering
        \includegraphics[width=\textwidth]{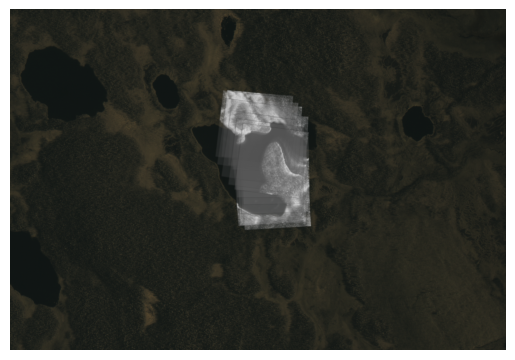}
        \caption{}
        \label{fig:PLSDAACZeroth}
    \end{subfigure}
    \hfill
    \begin{subfigure}{0.32\textwidth}
        \centering
        \includegraphics[width=\textwidth]{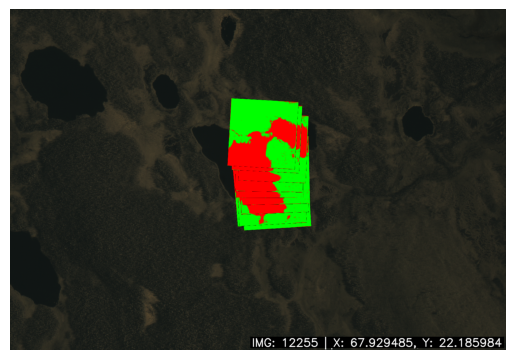}
        \caption{}
        \label{fig:PLSDAACCNN}
    \end{subfigure}
    \hfill
    \begin{subfigure}{0.32\textwidth}
        \centering
        \includegraphics[width=\textwidth]{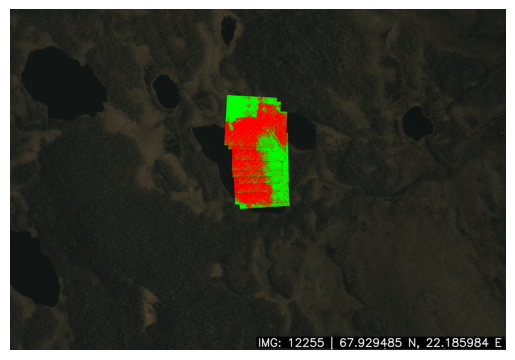}
        \caption{}
        \label{fig:PLSDAACEM}
    \end{subfigure}
    
    \caption{Autocorrelation of the CTIS image series 12255:12277. Panel (a) shows the 0th order of this series overlaid on the RGB image. Panel (b) shows the PLS-DA results from the CNN-based model overlaid on the RGB image, while panel (c) shows the PLS-DA results from the EM-based model overlaid on the RGB image. In (b) and (c) green classifies as land, while red classifies as water.}
    \label{fig:PLSDAAutocorrelation}
\end{figure}

Finally, Figure~\ref{fig:PLSDAAutocorrelation} presents a CTIS image series, where panel (a) displays the autocorrelation algorithm applied to the image series, overlaid on a corresponding RGB image of the same region. Panels (b) and (c) show the results of the PLS-DA models based on the respective reconstruction methods for the same image series.
For pixels that are covered by multiple CTIS images, the category visualized in panels (b) and (c) corresponds to the most frequent category present across the overlapping images.

\section{Conclusions}
In this study we have presented a first HAB test flight with a (prototype) snapshot hyperspectral camera, based on Computed Tomography Imaging Spectroscopy with a spectral range from 600-850 nm. 

The power system delivered sufficient power throughout the duration of the $>$5 hours flight  with 1.5 hours of ascent, 3.5 hours of float time and descent. 

Data was successfully acquired and stored and 3D hyperspectral datacubes were reconstructed during flight. The whole system was retrieved without any loss of data, and the camera also survived the impact with only minor damage to the outermost filter of the CTIS optics.

Based on the acquired data we have presented a proof-of-principle land cover classification, distinguishing land and water bodies along the flight track  based on the average reconstructed spectra.

A number of improvements are planned for the imaging system --- both to the optics and to the reconstruction algorithms --- and for the data link to allow for real time data transmission in order to deliver high quality videorate HSI imaging. 
%realize the potential of the HAB system to deliver high quality videorate HSI imaging for landcover classification. 
Our study has presented a promising first step in this direction and it represents the first application of this type of  CTIS camera outside of a lab environment. To our knowledge it is also the first time a CTIS camera has been succesfully integrated and deployed on a HAB for remote sensing applications.      

\section*{Declaration of generative AI and AI-assisted technologies in the writing process}
During the preparation of this work the authors used ChatGPT in order to improve language and readability. After using this tool/service, the authors reviewed and edited the content as needed and take full responsibility for the content of the publication.

\section*{Acknowledgements}
We gratefully acknowledge the REXUS/BEXUS programme - which is realised under a bilateral Agency Agreement between the German  Aerospace  Center (DLR) and the Swedish  National Space Agency (SNSA). The Swedish share of the payload has been made available to students from other European countries through a collaboration with the European Space Agency (ESA). EuroLaunch, a cooperation between the Swedish Space Corporation (SSC) and the Mobile Rocket Base (MORABA) of DLR, is responsible for the campaign management and operations of the launch vehicles. Experts from DLR, SSC, ZARM and ESA provide technical support to the student teams throughout the project. REXUS and BEXUS are launched from SSC, Esrange Space Center in northern Sweden.

Additionally, we acknowledge partial support from The Danish Industry Foundation, CenSec, SDU Climate Cluster, the T.B. Thrige Foundation, the Novo Nordisk Foundation, QTechnology and Newtec Engineering A/S, ICAPE Denmark A/S and AlmexA A/S.

\clearpage

\appendix
\section{Additional Data Analysis}
\label{sec:AppA}

\newcommand{\predictionmapping}[1]{
\begin{figure}[H]  % Ensure [H] is here
    \centering
    \begin{subfigure}[b]{0.44\textwidth}
        \centering
        \includegraphics[width=\textwidth]{CNN_rec_#1.png}
        \caption{Shows the spectral averaged CNN reconstruction of image #1}    
        \label{fig:cnn_rec_#1}
    \end{subfigure}
    \hfill
    \begin{subfigure}[b]{0.54\textwidth}  
        \centering 
        \raisebox{0.35cm}{\includegraphics[width=\textwidth]{Overlay_CNN#1.png}}
        \caption{Shows the CNN-based PLS-DA prediction map using 25 components, as a transparent overlay over (a)} 
        \label{fig:overlay_cnn_#1}
    \end{subfigure}
    
    \vskip\baselineskip
    
    \begin{subfigure}[b]{0.44\textwidth}   
        \centering 
        \includegraphics[width=\textwidth]{EM_rec_#1.png}
        \caption{Shows the spectral averaged EM reconstruction of image #1}    
        \label{fig:em_rec_#1}
    \end{subfigure}
    \hfill
    \begin{subfigure}[b]{0.54\textwidth}   
        \centering 
        \raisebox{0.35cm}{\includegraphics[width=\textwidth]{Overlay_EM#1.png}}
        \caption{Shows the EM-based PLS-DA prediction map using 15 components, as a transparent overlay over (c)}
        \label{fig:overlay_em_#1}
    \end{subfigure}
    
    \caption{Comparison of CTIS images with PLS-DA prediction maps from CNN and EM reconstruction respectively} 
    \label{fig:mean_and_std_#1}
\end{figure}
}

\predictionmapping{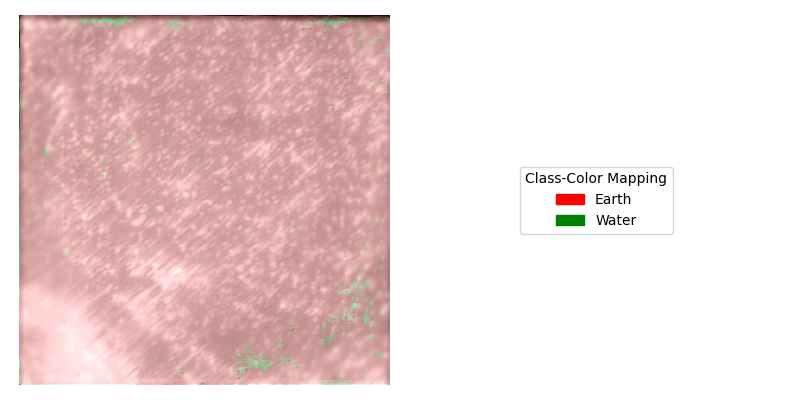}
\predictionmapping{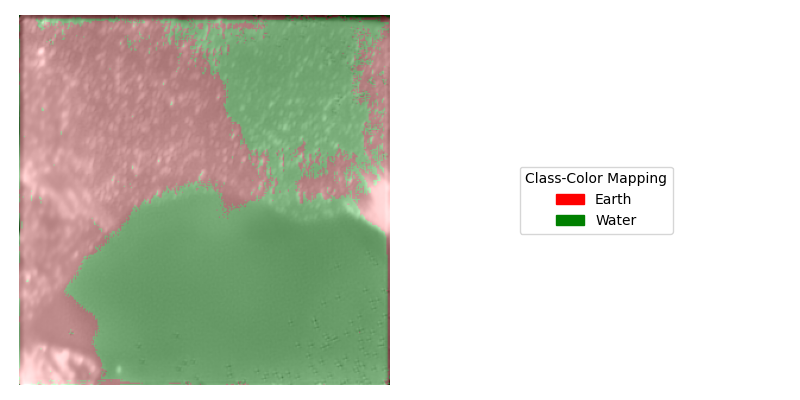}
\predictionmapping{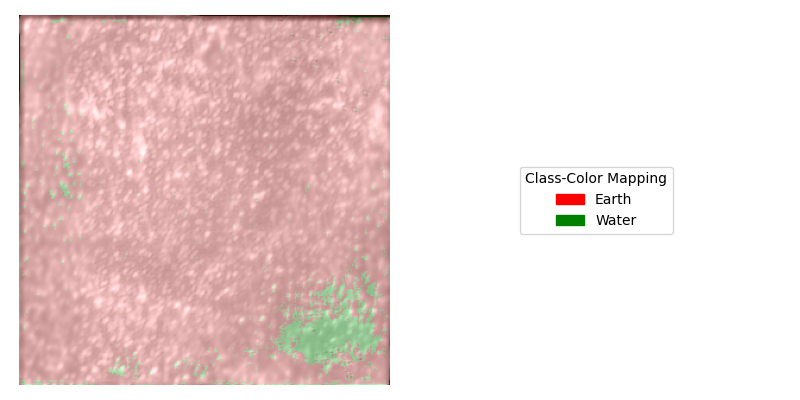}
\predictionmapping{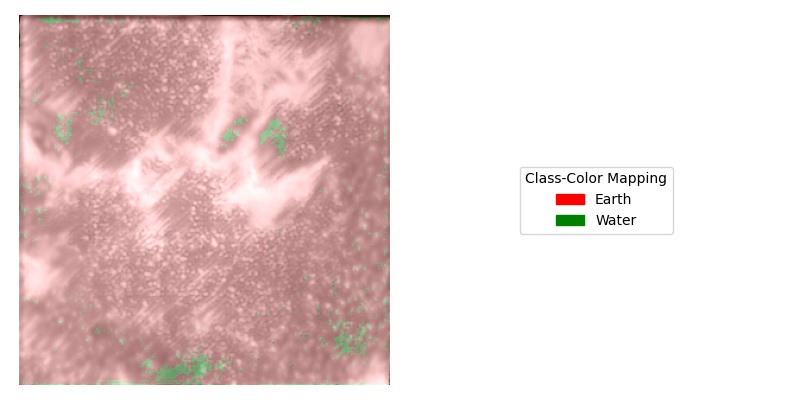}
\predictionmapping{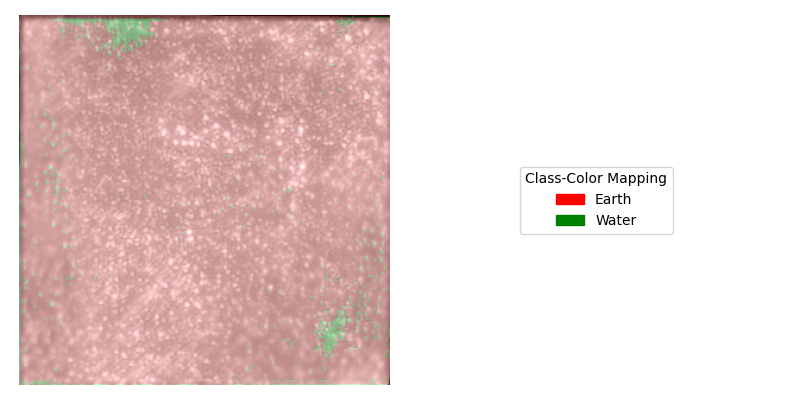}


\clearpage

\begin{thebibliography}{10}
\expandafter\ifx\csname url\endcsname\relax
  \def\url#1{\texttt{#1}}\fi
\expandafter\ifx\csname urlprefix\endcsname\relax\def\urlprefix{URL }\fi
\expandafter\ifx\csname href\endcsname\relax
  \def\href#1#2{#2} \def\path#1{#1}\fi

\bibitem{Floberghagen}
R.~Floberghagen, \href{https://atpi.eventsair.com/haps4esa2024/}{Why {HAPS} for {EO}}, HAPS4ESA (2024).
\newline\urlprefix\url{https://atpi.eventsair.com/haps4esa2024/}

\bibitem{Pankine}
A.~Pankine, Z.~Li, D.~Parsons, M.~Purucker, E.~Weinstock, W.~Wiscombe, K.~Nock, Stratospheric satellites for earth observations, Bulletin of The American Meteorological Society - BULL AMER METEOROL SOC 90 (08 2009).
\newblock \href {https://doi.org/10.1175/2009BAMS2624.1} {\path{doi:10.1175/2009BAMS2624.1}}.

\bibitem{Drinkwater}
M.~Drinkwater, \href{https://atpi.eventsair.com/haps4esa2024/}{Future{EO}}, HAPS4ESA (2019).
\newline\urlprefix\url{https://atpi.eventsair.com/haps4esa2024/}

\bibitem{AlexanderF}
A.~F.~H. Goetz, G.~Vane, J.~E. Solomon, B.~N. Rock, Imaging spectrometry for earth remote sensing, Science 228~(4704) (1985) 1147--1153.
\newblock \href {https://doi.org/10.1126/science.228.4704.1147} {\path{doi:10.1126/science.228.4704.1147}}.

\bibitem{boldrini2012hyperspectral}
B.~Boldrini, W.~Kessler, K.~Rebner, R.~W. Kessler, Hyperspectral imaging: a review of best practice, performance and pitfalls for in-line and on-line applications, Journal of near infrared spectroscopy 20~(5) (2012) 483--508.

\bibitem{descour_computed-tomography_1995}
M.~Descour, E.~Dereniak, Computed-tomography imaging spectrometer: experimental calibration and reconstruction results, Appl. Opt. 34~(22) (1995) 4817--4826.
\newblock \href {https://doi.org/10.1364/AO.34.004817} {\path{doi:10.1364/AO.34.004817}}.

\bibitem{Th}
T.~V. Bulygin, G.~N. Vishnyakov, \href{https://doi.org/10.1117/12.131904}{{Spectrotomography: a new method of obtaining spectrograms of two-dimensional objects}}, in: G.~G. Levin (Ed.), Analytical Methods for Optical Tomography, Vol. 1843, International Society for Optics and Photonics, SPIE, 1992, pp. 315 -- 322.
\newline\urlprefix\url{https://doi.org/10.1117/12.131904}

\bibitem{Okamoto:91}
T.~Okamoto, I.~Yamaguchi, Simultaneous acquisition of spectral image information, Opt. Lett. 16~(16) (1991) 1277--1279.
\newblock \href {https://doi.org/10.1364/OL.16.001277} {\path{doi:10.1364/OL.16.001277}}.

\bibitem{huang_application_2022}
W.-C. Huang, M.~S. Peters, M.~J. Ahlebaek, M.~T. Frandsen, R.~L. Eriksen, B.~Jørgensen, The application of convolutional neural networks for tomographic reconstruction of hyperspectral images, Displays (2022) 102218\href {https://doi.org/10.1016/j.displa.2022.102218} {\path{doi:10.1016/j.displa.2022.102218}}.

\bibitem{ahlebaek_hybrid_2023}
M.~Ahlebæk, M.~Peters, W.-C. Huang, M.~Frandsen, R.~Eriksen, B.~Jørgensen, The hybrid approach—convolutional neural networks and expectation maximisation algorithm—for tomographic reconstruction of hyperspectral images, J. Spectral Imaging (2023) a1\href {https://doi.org/10.1255/jsi.2023.a1} {\path{doi:10.1255/jsi.2023.a1}}.

\bibitem{10.1117/12.2621128}
M.~S. Peters, R.~L. Eriksen, B.~J{\o}rgensen, \href{https://doi.org/10.1117/12.2621128}{{High-resolution snapshot hyperspectral computed tomography imaging spectrometer: real-world applications}}, in: M.~P. Georges, G.~Popescu, N.~Verrier (Eds.), Unconventional Optical Imaging III, Vol. 12136, International Society for Optics and Photonics, SPIE, 2022, p. 121360R.
\newblock \href {https://doi.org/10.1117/12.2621128} {\path{doi:10.1117/12.2621128}}.
\newline\urlprefix\url{https://doi.org/10.1117/12.2621128}

\bibitem{PETERS2025126017}
M.~S. Peters, M.~J. Ahlebæk, M.~T. Frandsen, B.~Jørgensen, C.~H. Jessen, A.~K. Carlsen, M.~S. Andersen, W.-C. Huang, R.~L. Eriksen, Investigating the applicability of a snapshot computed tomography imaging spectrometer for the prediction of °brix and ph of grapes, Spectrochimica Acta Part A: Molecular and Biomolecular Spectroscopy 336 (2025) 126017.
\newblock \href {https://doi.org/https://doi.org/10.1016/j.saa.2025.126017} {\path{doi:https://doi.org/10.1016/j.saa.2025.126017}}.

\bibitem{MAYRA2021112322}
J.~Mäyrä, S.~Keski-Saari, S.~Kivinen, T.~Tanhuanpää, P.~Hurskainen, P.~Kullberg, L.~Poikolainen, A.~Viinikka, S.~Tuominen, T.~Kumpula, P.~Vihervaara, Tree species classification from airborne hyperspectral and lidar data using 3d convolutional neural networks, Remote Sensing of Environment 256 (2021) 112322.
\newblock \href {https://doi.org/https://doi.org/10.1016/j.rse.2021.112322} {\path{doi:https://doi.org/10.1016/j.rse.2021.112322}}.

\bibitem{rs8060445}
L.~Ballanti, L.~Blesius, E.~Hines, B.~Kruse, Tree species classification using hyperspectral imagery: A comparison of two classifiers, Remote Sensing 8~(6) (2016).
\newblock \href {https://doi.org/10.3390/rs8060445} {\path{doi:10.3390/rs8060445}}.

\bibitem{Masaitis}
G.~Masaitis, G.~Mozgeris, Some peculiarities of laboratory measured hyperspectral reflectance characteristics of scots pine and norway spruce needles, Research for Rural Development 2 (2012) 25--32.

\bibitem{nelson2022}
P.~R. Nelson, A.~J. Maguire, Z.~Pierrat, E.~L. Orcutt, D.~Yang, S.~Serbin, G.~V. Frost, M.~J. Macander, T.~S. Magney, D.~R. Thompson, J.~A. Wang, S.~F. Oberbauer, S.~V. Zesati, S.~J. Davidson, H.~E. Epstein, S.~Unger, P.~K.~E. Campbell, N.~Carmon, M.~Velez-Reyes, K.~F. Huemmrich, Remote sensing of tundra ecosystems using high spectral resolution reflectance: Opportunities and challenges, Journal of Geophysical Research: Biogeosciences 127~(2) (2022) e2021JG006697, e2021JG006697 2021JG006697.
\newblock \href {https://doi.org/https://doi.org/10.1029/2021JG006697} {\path{doi:https://doi.org/10.1029/2021JG006697}}.

\bibitem{Delaure}
B.~Delauré, J.~Blommaert, S.~Livens, D.~Nuyts, G.~Strackx, T.~Van~Achteren, High resolution hyperspectral imaging from haps, hAPS4ESA (10 2017).

\bibitem{navarro_novel_2022}
P.~J. Navarro, L.~Miller, M.~V. Díaz-Galián, A.~Gila-Navarro, D.~J. Aguila, M.~Egea-Cortines, A novel ground truth multispectral image dataset with weight, anthocyanins, and {Brix} index measures of grape berries tested for its utility in machine learning pipelines, GigaScience 11 (2022) giac052.
\newblock \href {https://doi.org/10.1093/gigascience/giac052} {\path{doi:10.1093/gigascience/giac052}}.

\bibitem{peters_high-resolution_2022}
M.~S. Peters, R.~L. Eriksen, B.~Jørgensen, High-resolution snapshot hyperspectral computed tomography imaging spectrometer: real-world applications, in: Unconventional {Optical} {Imaging} {III}, Vol. 12136, SPIE, 2022, pp. 211--216.
\newblock \href {https://doi.org/10.1117/12.2621128} {\path{doi:10.1117/12.2621128}}.

\bibitem{mortonarboretum_norwayspruce}
{The Morton Arboretum}, Norway spruce (picea abies), accessed: [date] (n.d.).

\bibitem{tree_cluster}
B.~R. Parizek, K.~Christianson, R.~B. Alley, D.~Voytenko, I.~Vaňková, T.~H. Dixon, R.~T. Walker, D.~M. Holland, {Ice-cliff failure via retrogressive slumping}, Geology 47~(5) (2019) 449--452.
\newblock \href {https://doi.org/10.1130/G45880.1} {\path{doi:10.1130/G45880.1}}.

\bibitem{white_accelerating_2020}
L.~White, W.~B. Bell, R.~Haygood, Accelerating computed tomographic imaging spectrometer reconstruction using a parallel algorithm exploiting spatial shift-invariance, Opt. Eng. 59~(05) (2020) 1.
\newblock \href {https://doi.org/10.1117/1.OE.59.5.055110} {\path{doi:10.1117/1.OE.59.5.055110}}.

\bibitem{douarre_ctis-net_2021}
C.~Douarre, C.~Crispim-Junior, A.~Gelibert, G.~Germain, L.~Tougne, D.~Rousseau, {CTIS}-net: A neural network architecture for compressed learning based on computed tomography imaging spectrometers, {IEEE} Trans. Comput. Imaging 7 (2021) 572--583.
\newblock \href {https://doi.org/10.1109/TCI.2021.3083215} {\path{doi:10.1109/TCI.2021.3083215}}.

\bibitem{mel2022joint}
M.~Mel, A.~Gatto, P.~Zanuttigh, Joint reconstruction and super resolution of hyper-spectral ctis images., in: BMVC, 2022, p. 1063.

\bibitem{Wu}
L.~Wu, W.~Cai, Ctis-gan: computed tomography imaging spectrometry based on a generative adversarial network, Applied Optics 62 (03 2023).
\newblock \href {https://doi.org/10.1364/AO.478230} {\path{doi:10.1364/AO.478230}}.

\bibitem{Zimmermann}
M.~Zimmermann, S.~Amann, M.~Mel, T.~Haist, A.~Gatto, Deep learning-based hyperspectral image reconstruction from emulated and real computed tomography imaging spectrometer data, Optical Engineering 61 (05 2022).
\newblock \href {https://doi.org/10.1117/1.OE.61.5.053103} {\path{doi:10.1117/1.OE.61.5.053103}}.

\end{thebibliography}
\end{document}